\documentclass[twocolumn,a4paper,10pt]{article} 
\usepackage{epsfig}
\usepackage{float}
\usepackage{subfigure}
\usepackage{pstricks}
\begin{document}
\pagestyle{plain}
%
%

\title{
{\Large \bf Heavy quarkonium production in hadronic collisions\footnote{To appear in the proceedings of the JJC2001 --Journ\'ees Jeunes Chercheurs--, La~Hume, France, 10-14 December 2001.}}}
 
\author{
{\bf Jean-Philippe LANSBERG\footnote{JPH.Lansberg@ulg.ac.be}} \\ 
Groupe de Physique Th\'eorique Fondamentale\\
Universit\'e de Li\`ege\\
Institut de Physique, B\^at. B5 \\
B-4000 Li\`ege 1 BELGIUM 
}

\date{}

\maketitle

%
\begin{center}
{\bf 
Abstract
}\\
\end{center}

Different theoretical models which attempt to describe hadronic production of 
heavy quarkonia are reviewed. Firstly, we consider the Color Singlet Model and 
point out the large discrepancies between the theoretical predictions and the 
results from the Tevatron detectors. Then some other models are introduced, 
quickly discussed and confronted with experimental results. Finally, we 
suggest possible ways to understand the source of the remaining discrepancies.

\vspace{-0.5cm}

%
\section{Introduction}
%
Since the discovery of the $J/\psi$ and of some higher states of charmonia, the 
calculation of their production rates has been performed. In the early 80's, there were 
two leading models for the description of the data, namely the Color Singlet 
Model (CSM)~\cite{CSM} and the Color Evaporation Model (CEM)~\cite{CEM}. Each were 
based on simple but justifiable assumptions that we shall discuss later on. 
 
Until the middle of the 90's, the experimental results were all in good agreement 
with these two models in any type of production but the CSM was preferred 
because of a seemingly more solid theoretical foundation. This agreement was 
largely due to the fact that direct production was not observed, so that the 
$J/\psi$ would result from a variety of cascading decays. Nevertheless, in 1995, the CDF 
collaboration showed large discrepancies between the rates of $\psi(2S)$
 predicted by the CSM and experimental results for
direct production in high energy $p \bar p$ collisions~\cite{CDF7997a}. The same 
occurred for the $J/\psi$ state in 1997 when they achieved the isolation of the direct 
contribution~\cite{CDF7997b}.

This discovery gave the CEM a second life despite its weaker foundation compared
 to the CSM. Besides that, some 
other models or mechanisms were proposed to solve the problem. The first was 
the Color Octet Mechanism (initially introduced for high-$p_T$ fragmentation). 
Its key-point is that the bound state can be produced in a colored state and 
then bleached into a singlet state by non-perturbative processes, these last effects
 being mathematically given by non-perturbative matrix elements~\cite{BBL1}.

Later, this mechanism was included in a more general formalism, non-relativistic
 QCD (NRQCD). The latter is based on a systematic expansion in the coupling 
constant and the quark velocity in the bound state, which for heavy quarkonia is supposed
to be much less than the speed of light.

The theoretical predictions based on NRQCD account well for all the available data 
from hadron colliders and more or less satisfactorily for data from {\it e-p} colliders ({\it e.g.} 
HERA). The only discrepancies come from polarization measurements, where NRQCD predicts
a transverse polarization, the data clearly do not show signs of any polarization
\cite{polarization}. 
 
Another model was proposed by P. Hoyer and S. Peign\'e \cite{hoyer}. The basic idea 
is that the heavy-quark pair can undergo a perturbative interaction with the comoving 
color-field produced by the initial hadronic collision. Introducing a new 
variable which parameterizes this interaction, it is able to reproduce some features
of the data which are not described by the CSM, as well as polarization measurements.

\vspace{-0.5cm}

%
\section{The Color Singlet Model}
%

\subsection{The model}

This model is based on several approximations or postulates:
\begin{itemize}
\item Decomposing the quarkonium production in two steps, first the creation of 
the 2 heavy quarks ($Q\ \& \ \bar Q$) and then the binding of these two quarks 
forming the meson, one {\it postulates the factorization} of these two processes.
\item As the scale of the first process is approximately $M^2+p_T^2$ one  
considers it as a {\it   perturbative} one. One can thus calculate its cross section 
with classical Feynman-diagram methods.
\item Considering only bound states of heavy quarks, their velocity
 in the meson must be small. One therefore supposes that the meson is created 
with its 2 constituent quarks {\it   at rest} in the meson frame. This 
is {\it the static approximation.}
\item One finally assumes that the color and the spin of the pair $Q\bar Q$ don't 
change during the binding. As physical states are colorless, one requires the pair 
to be produced in a {\it   color singlet state}. 
\end{itemize}

In high-energy hadronic collisions, the leading contribution comes from a gluon 
fusion process. Using a meson production vertex function with the required 
tensorial structure and prescriptions relative to  the propagators,  one has
six diagrams for the $^3S_1$ states,
\begin{figure}[H]
\centering
\psfig{figure=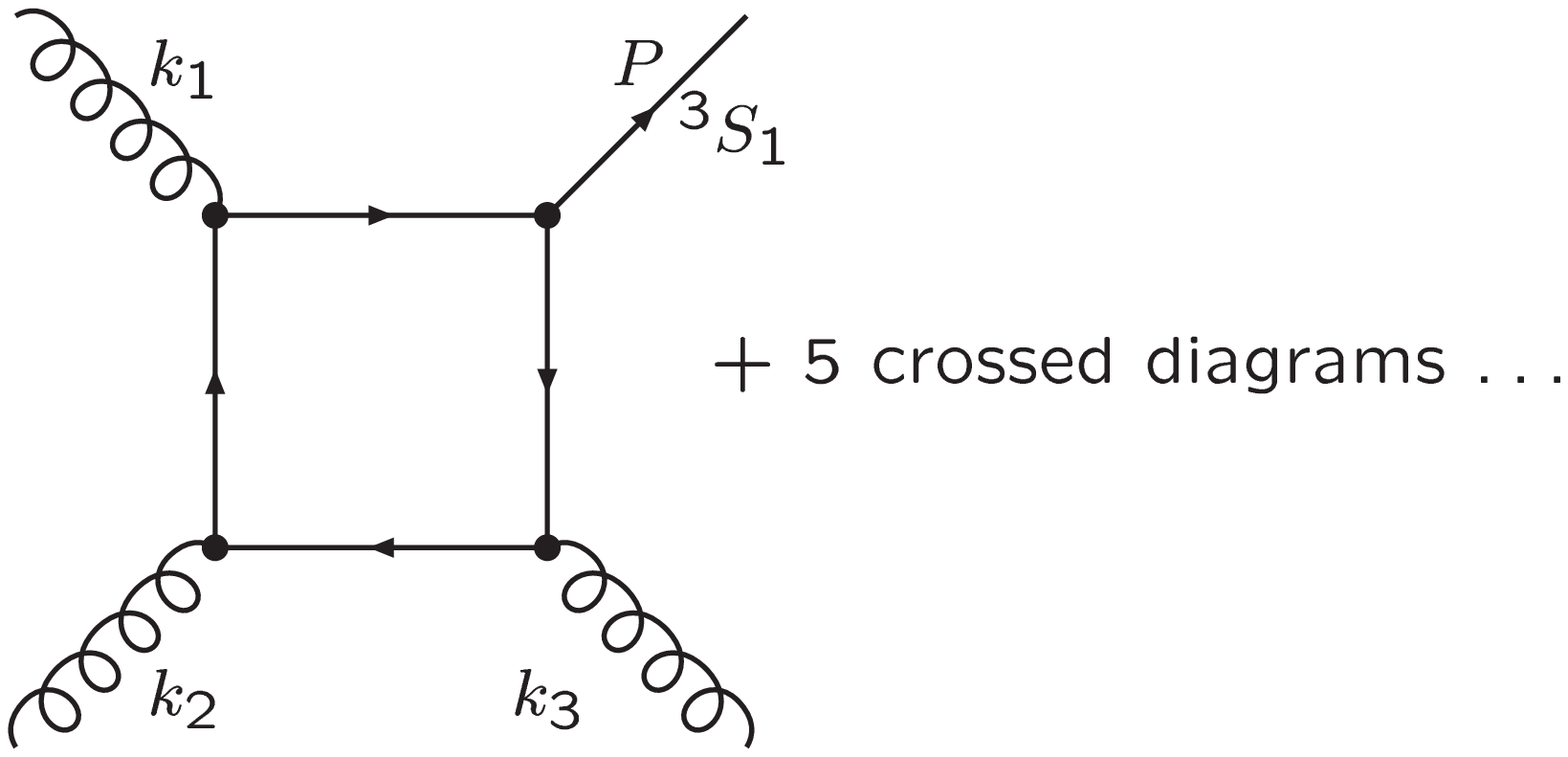,width=6cm}
\end{figure} 
which give
\begin{eqnarray}\label{eq:sigma}
&\ & \hspace{-0.5cm} \frac{d\sigma}{d\hat t}= \frac{20\pi^2 M \alpha_s^3|\psi(0)|^2}{9\hat s^2} \\ 
&\times & \frac{\left[\hat s^2
(\hat s-M^2)^2\right]+\left[\hat t^2(\hat t-M^2)^2\right]+\left[\hat u^2(\hat u-M^2)^2\right]}
{(\hat s-M^2)^2(\hat t-M^2)^2(\hat u-M^2)^2} \nonumber
\end{eqnarray}

\subsection{Comparison with data}
\begin{figure}[h]
\centering
\psfig{figure=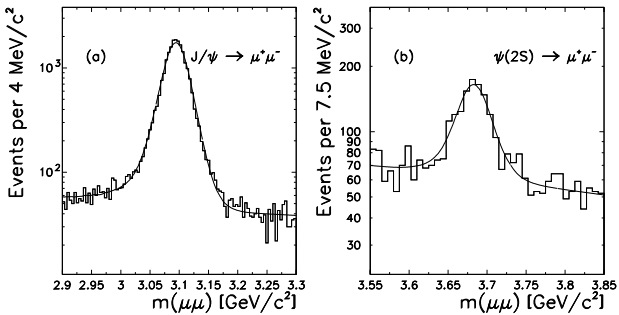,width=9cm}
\caption{Resonances due to $J/\psi$ (a) and to $\psi(2S)$ (b)~\cite{CDF7997a}.}
\label{fig:reson}
\end{figure}

\begin{figure}[h]
\centering
\psfig{figure=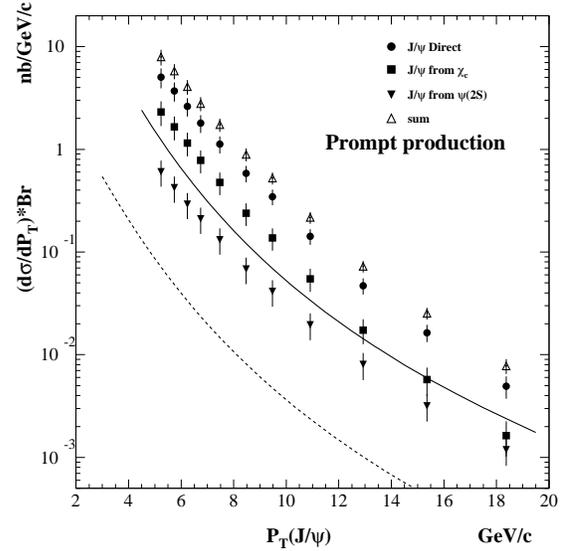,width=8cm}
\caption{$\frac{d\sigma}{dP_T}{\cal B}$ as a function of $P_T $ for $J/\psi$~\cite{CDF7997b}.}
\label{fig:dsdptjpsi}
\end{figure}

\begin{figure}[h]
\centering
\psfig{figure=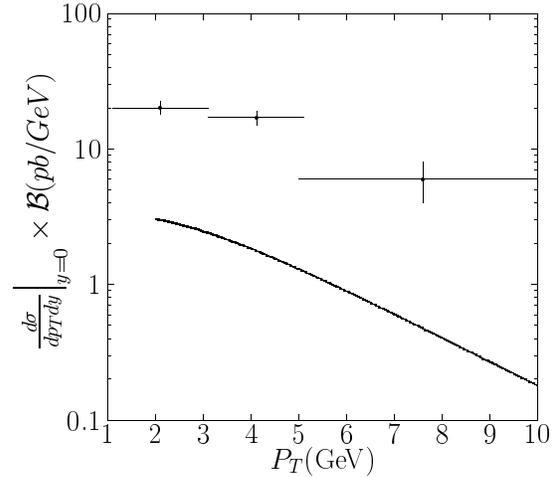,width=6.7cm}
\caption{$\frac{d\sigma}{dP_T}{\cal B}$  for {\bf direct} $\Upsilon(3S)$~\cite{CDF7595}.}
\label{fig:dsdptups3s}
\end{figure}

Given their quite large branching ratio into dileptons, the best 
way to detect (heavy)quarkonia is to focus on muon pairs and to plot their 
invariant mass distribution (cf. Figure~\ref{fig:reson} (a)\&(b)). For instance, 
in the mass region of the $J/\psi$, 
the distribution shows a maximum at the precise value of $m_{J/\psi}$. 
Constrained fits on this distribution for different values of kinematical 
parameters then provide us with differential cross sections relative to 
these parameters (cf. Figure~\ref{fig:dsdptjpsi} and~\ref{fig:dsdptups3s} 
for $\frac{d\sigma}{dP_T}$).

In order to select the type of production, some other constraints can be imposed. 
Non-Prompt production (coming from $b$ quark decay) is rejected by the detection of 
a secondary vertex. The prompt but non-direct production (coming from  
radiative decay --cf. Figure~\ref{fig:dsdptjpsi}: squares and plain triangles--) 
is rejected by detecting the photon emitted during the decay.

Figures~\ref{fig:dsdptjpsi} and \ref{fig:dsdptups3s} show the curve obtained 
with the CSM (dotted curve for Figure~\ref{fig:dsdptjpsi}) and the measurements by the CDF 
Collaboration. In the case of the 
$J/\psi$, the discrepancies are more or less a factor of 30, for the 
$\psi(2S)$ (not shown) they reach 60 and for the $\Upsilon(nS)$ (only the $\Upsilon(3S)$ 
plot is shown in Figure~\ref{fig:dsdptups3s}) the factor is
10. 

It is therefore evident that the CSM totally fails to reproduce the data. The 
same thing happens with D$\emptyset$ results. An experimental problem is thus unlikely. 
Another important feature to 
note is that the electroproduction data (from HERA detectors) are up to now still 
in good agreement with CSM predictions.

%
\section{The Color Evaporation Model}
%

This model is based on the fact that $\alpha_{\rm strong} >1$ for long range 
interactions 
or low 4-momentum transfers. As a consequence, the probability ${\cal P}$ that a quark pair 
undergoes many quantum fluctuations during non-perturbative interactions with 
surrounding hadronic matter is big. Schematically,\\

\begin{figure}[H]
\centering
\psfig{figure=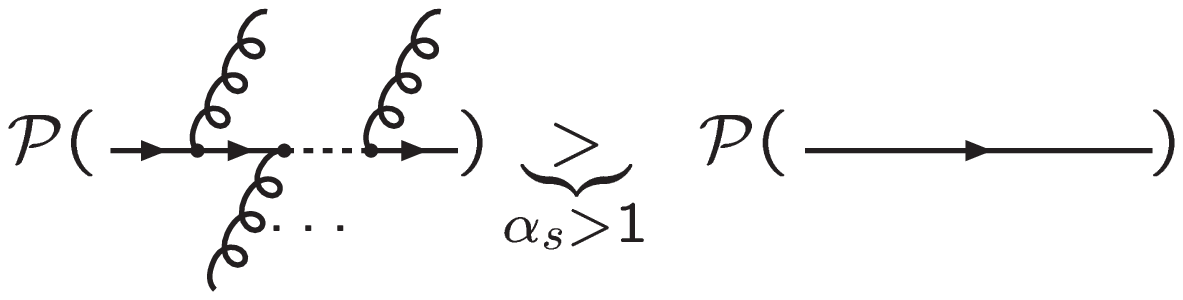,width=6cm}
\label{fig:CEM0}
\end{figure} 

The asymptotic state (the meson) is random, still being colorless. Thus the 
probability to produce the different quarkonia states of a given family 
is supposed equal, or almost equal.

Mathematically, this gives
\begin{equation}
\sigma(^{2S+1}L_J)= \frac{F}{9}\int_{2m_{c,b}}^{2m_{D,B}}dm \frac{d\sigma_{c\bar c,b\bar b}}{dm},
\end{equation}
where the {\it natural} value of $F$ is the inverse of the number of quarkonia
between the threshold  $2m_{c,b}$ and $2m_{D,B}$.

The new feature compared to the CSM is that the leading contribution at 
low $P_T$ consists in the following process
\begin{figure}[H]
\centering 
\psfig{figure=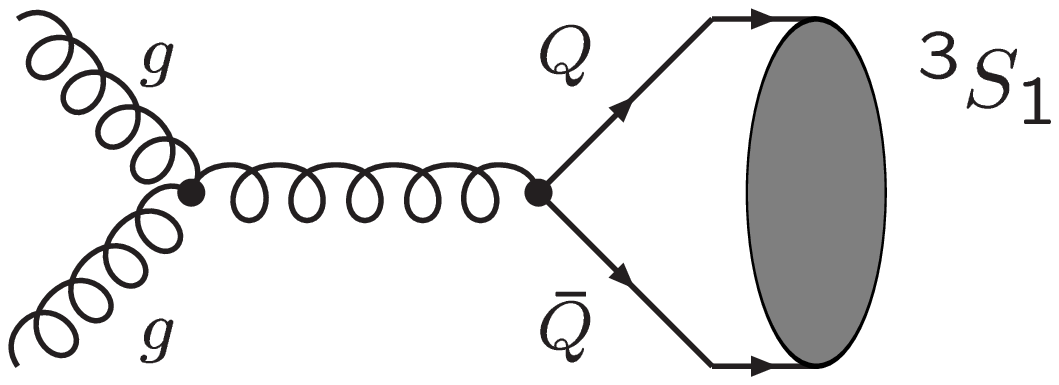,width=5cm}
\label{fig:CEM1}
\end{figure} 

As already mentioned, this model is in good agreement with experimental data. 
For instance, it reproduces quite well the energy dependence of the cross section 
as well as its polarization.

Nevertheless this model raises several remarks:

\begin{itemize}
\item It is very phenomenological. 
\item $F$ is in fact a free parameter, its fitted value seems to depend 
on the kinematics.
\item In order to obtain $\frac{d\sigma}{dP_T}$, one is tempted to introduce NLO 
contributions. This is by construction of the model normally included in $F$, and 
hence one may be double-counting.
\end{itemize}

%
\section{The Model of Hoyer \& Peign\'e.}
%

\begin{figure}[h]
\centering
\psfig{figure=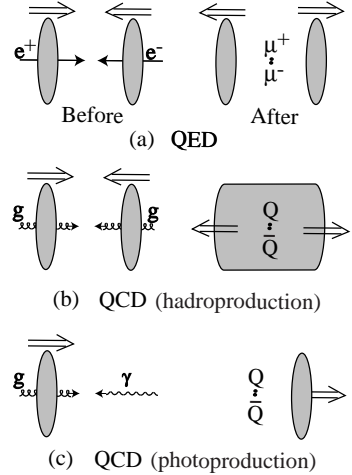,width=4.5cm,angle=-0.5}
\caption{Due to self-interaction of gluons, the gluon field produced by Bremsstrahlung
stays at rest in the rest frame of the quark pair. }
\label{fig:hoyer}
\end{figure}

This model describes the production of quarkonium through Hard Comover Scattering. Some 
features of other related processes, {\it e.g.} open production, 
indeed suggest that there 
exists a comoving gluonic field produced by the hadronic collisions. One can convince 
oneself that such field doesn't exist in general in QED, and in particular in 
leptoproduction of quarkonium (cf. Figure~\ref{fig:hoyer}). This could explain why 
the problem does not occur in {\it e-p} collisions as in $p\bar p$ collisions. Moreover, 
the re-interaction has to be perturbative because the heavy quark symmetry predicts that 
non-perturbative interactions will not change spin. In this sense the CEM contradicts 
this model and the heavy quark symmetry.

The authors of this model can thus reproduce the experimental data by fitting the variable
which parameterizes the scattering with some assumptions on the topology of the 
field and on its polarization.

The main success of the model is its physical content, its ability to explain
other features than the simple production of quarkonium-- for instance $J/\psi$ 
suppression in hadronic matter-- and, as hoped, the results directly linked to 
production cross sections of vector mesons.

%
\section{Analysis of the theoretical uncertainties.}
%

In order to get a first idea of what could be the source of such discrepancies,
we have undertaken an analysis of the 
theoretical uncertainties arising in the CSM. The first source we've considered 
is related to the wave function at the origin, which enters 
directly the expression of the cross section (\ref{eq:sigma}). 

Its value is in fact extracted from the leptonic decay width, which writes 

\begin{equation}
\Gamma(^3S_1 \to \ell \bar\ell)= \frac{64 \pi}{9} \frac{\alpha^2 e^2_Q\psi^2(0)}{M^2}.
\end{equation}

We thus find that the error on  $\Gamma_{\mu\mu}$ introduces an error of at least 
$10 \%$  on the cross-section.

\begin{center}
\begin{tabular}{|c|c|c|}
\hline Meson & $|\psi(0)|^2 \pm \sigma_{|\psi(0)|^2}$ & Relative error  \\
\hline \hline 
$J/\psi $        &  $0.041 \pm 0.0042$ GeV$^3$ & $10 \%$	\\
$\psi(2S) $      &  $0.024 \pm 0.0025$ GeV$^3$ & $10 \%$	\\
$\Upsilon(1S)$   &  $0.397 \pm 0.015  $  GeV$^3$ & $4 \%$ 	\\
$\Upsilon(2S)$   &  $0.192 \pm 0.030  $  GeV$^3$ & $16 \%$	\\
$\Upsilon(3S)$   &  $0.173 \pm 0.032  $  GeV$^3$ & $18 \%$ \\
\hline
\end{tabular}
\end{center}

Then we analyzed the uncertainties due to the parton distribution functions (pdf). 
There exist two ways by which the pdf's can introduce errors in the cross section. Firstly the 
value of $\alpha_S(Q^2)$ attached with the pdf and secondly the pdf's 
themselves for the hadronic cross section.   

We found that the overall factor resulting from theses 3 sources of uncertainties is 
about 2-3 and roughly constant for different values of $P_T$.

\begin{figure}[H]
\psfig{figure=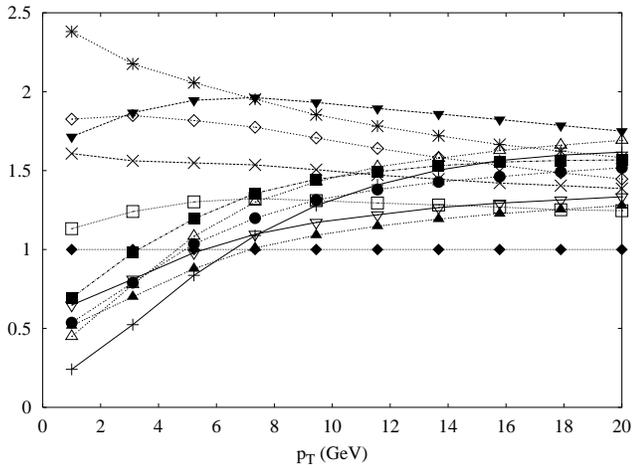,width=9cm}
\caption{Ratio of the cross sections obtained with various pdf's~\cite{pdflib} to the 
cross section obtained with MRS(G) 2-94.}
\label{fig:deviation}
\end{figure}

%
\section{Summary and Outlooks}
%
In this brief review, we saw that the CSM was unable to reproduce the experimental 
data, especially for charmonia, even if it is based on sensible approximations 
and postulates. This feature was discovered nearly ten years ago for the 
$\psi'$. Even if NRQCD was believed to be the appropriate answer, 
according to the recent data on polarization, its efficiency is now 
arguable. Other models, CEM and re-scattering model, are efficient but 
we need more tests to reinforce their credibility. 

In this context, the evaluation of the non-static contribution of the 
CSM could be one of the solutions. It could open new paths for the 
understanding of this 
serious problem and could help for a better understanding of relativistic 
wave functions and of gauge invariance in bound states description.

%
%
%
\vspace*{0.5cm}
\begin{center}
{\large\bf Acknowledgments}
\end{center}

 I gratefully thank J-R.~Cudell for a careful reading of this report
and the organizers for the time spent for organizing 
this meeting. This work is supported by the Institut Interuniversitaire 
des Sciences Nucl\'eaires (IISN).

%
%

\end{document}